\documentclass[aps,twocolumn,showpacs]{revtex4}
\usepackage{dcolumn}
\usepackage{graphicx}
\usepackage{amsmath}
\usepackage{verbatim}
\usepackage{amssymb}
\usepackage{psfrag}
\usepackage{bm}
\usepackage{braket}
\usepackage{wrapfig}
\usepackage{subfigure}
\usepackage{makeidx}
\usepackage{bm}
\usepackage{epsf}
\usepackage{multirow}
\usepackage[colorlinks,linkcolor=blue,urlcolor=blue,citecolor=blue]{hyperref}

\DeclareMathOperator{\sech}{sech}
\begin{document}
	\title{One family  of dark-bright solitons with striking width differences}
	\author{Ning Mao}
	\author{Li-Chen Zhao}\email{zhaolichen3@nwu.edu.cn}
	\affiliation{$^{1}$School of Physics, Northwest University, Xi'an 710127, China}
	\affiliation{$^{2}$Shaanxi Key Laboratory for Theoretical Physics Frontiers, Xi'an 710127, China}
	\affiliation{$^{3}$NSFC-SPTP Peng Huanwu Center for Fundamental Theory, Xi'an 710127, China}

	%%%%%%%%%%%%%%%%%%%%%%%%%%%%%%%%%%%%%%%%%%%%%%%%%
	\date{\today}
	\begin{abstract}
		Most of previously reported dark-bright solitons admit identical width for the two components in both theoretical and experimental studies.  We report dark-bright solitons can admit strikingly different widths, and  derive a family of analytical solutions for them by Lagrangian variational method. The existence regimes for these solitons  become much more widespread in the space of nonlinear parameters, than the ones for the previously known dark-bright solitons with identical width. Our analysis indicates that the effective quantum wells are quite different in the two components, in sharp contrast to the ones for all previously known vector solitons. Especially, the particle number of bright soliton can be used to control the generation of dark-bright solitons with varied ratios of solitons' widths. Based on the current experimental technologies, we propose an experimental scheme for observing  these novel dark-bright solitons. The results suggest that abundant vector solitons with difference widths exist in multi-components coupled systems,  and would inspire experiments to observe them in nonlinear optical fibers, Bose-Einstein condensates, and other nonlinear coupled systems.
	\end{abstract}
	\maketitle
	
\section{introduction}
Vector solitons capture significant attention in both experimental and theoretical research in Bose-Einstein condensates \cite{SoliPhy,BEC} due to their more abundant dynamics compared scalar soliton \cite{OS,cs1,cs2,cs3,cs4,cs5,cs6,cs7,cs8}, such as dark-dark, dark-bright, magnetic solitons, and others \cite{vs1,vs2,vs3,vs4,vs5,vs6}.  Many efforts have been paid to find the  analytical solutions for them, due to their beauty and convenience in unveiling the underlying physics \cite{AS1,AS2,Manakov,Darb1,Darb2,Darb3,inverse,Hirota,Jo,LVM,KDS}. We note that all those vector soliton solutions possess identical width for different components. However, there are no physical laws dictating that the solitons in different components must have the same width. These characters motivate us to discuss whether the vector solitons can admit different widths.

Among the vector solitons, dark-bright solitons has been paid more attention due to that the two components naturally admits distinctive eigenmodes for them \cite{cs3,cs4,DBO,multi-dark-bright,DBT}. The dark-bright solitons were used to discuss some striking motion \cite{DBBC,Spin AC,AC Oscillation} induced by the interplay of positive mass of a bright soliton and negative mass of a dark soliton. Moreover, they are much easier to be realized in experiments, with the aid of spatial-dependent Rabi coupling and phase imprinting techniques \cite{cs3}. Our recent studies on vector soliton solution indicate that dark-bright soliton solution is also helpful for deriving other vector soliton solutions \cite{LVM}. Therefore, we focus on dark-bright soliton solution as an example to address the above question.

In the paper, we derive a family of analytical solutions for dark-bright soliton with different widths by Lagrangian variational method \cite{VM1,VM2,VM3,VM4,VM5,VM6}. We show that its existence regimes in the space of nonlinear parameters is more extensive than the previously known exact solution with identical width \cite{Jo,DBO,LVM,Spin AC}. Since a nonlinear Schr\"{o}diner equation can be mapped into a linear Schr\"{o}diner equation with a corresponding quantum well \cite{qw1,qw2,qw3}, we analyze the effective quantum wells of the novel soliton. Our analysis reveals that the quantum wells are quite different in the two components, in sharp contrast to the ones for all previously known vector solitons. This difference in quantum wells provides insight into understanding why the coupled system admits vector solitons with different widths. Especially, we demonstrate that the ratio of widths between the two components can be controlled by the particle number of bright soliton while keeping the  velocity fixed. Based on current experimental technologies and our results, we propose an experimental scheme for observing these novel dark-bright solitons. These efforts would motivate experiments to observe the novel vector solitons in a coupled nonlinear system.

This paper is organized as follows. In Sec. \ref{ii}, we provide the existence conditions of different widths dark-bright solitons obtained from Lagrangian variational method, and we suggest a possible reason for the existence of these novel solitons in coupled system. Then we demonstrate that the ratio of widths between two components can be controlled by the particle number of bright soliton while keeping the velocity fixed. In Sec. \ref{iii}, considering a quasi-one-dimensional two-component Bose-Einstein condensates
with transition effect,  we successfully generate the dark-bright solitons with different ratios of widths in  numerical simulations, manipulating the transition time to control the particle number. Additionally, we propose an experimental scheme to generate the different widths dark-bright soliton in Bose-Einstein condensates. Finally, our conclusions and discussions are given in Sec. \ref{iv}.
	
\section{The characteristics of different widths dark-bright soliton}\label{ii}
\subsection{The existence of the solitons }
\begin{figure}[t]
	\centering
	\includegraphics[width=86mm]{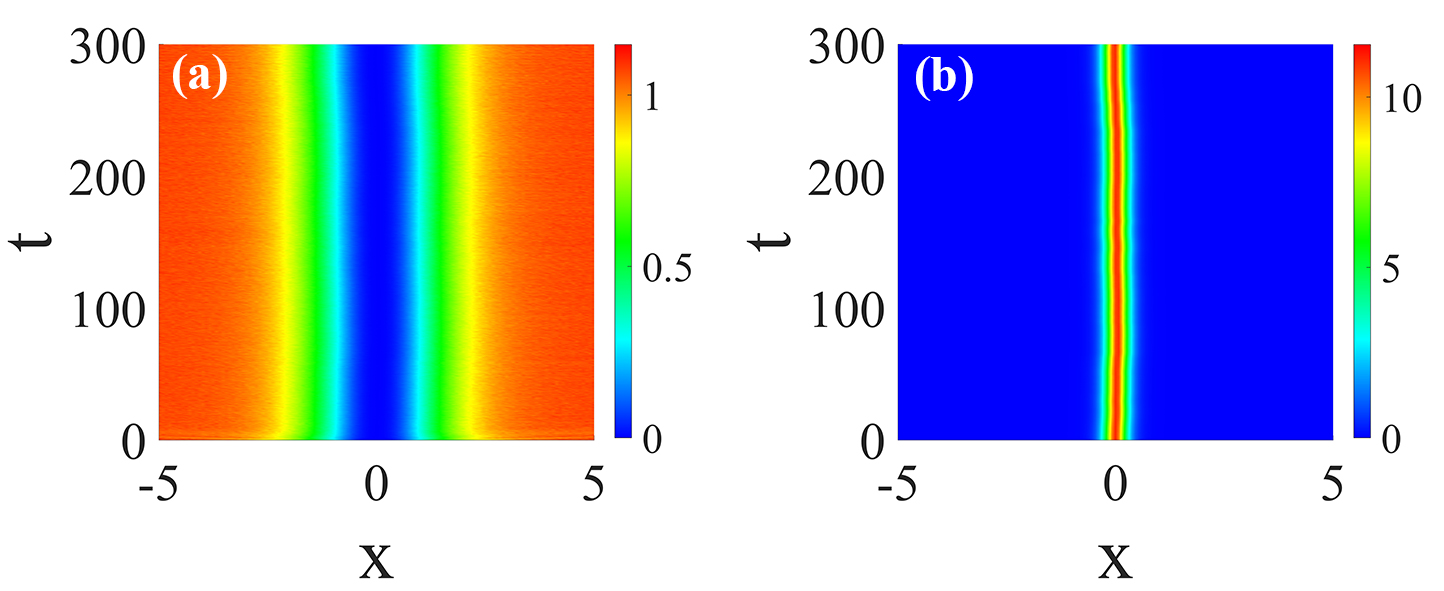}
	\caption{(a)(b) The numerical evolution of different widths  dark-bright (DWDB) soliton. The initial state is given by Lagrangian method and introducing a  $3\%$random noise. It is seen that the soliton still remains stable for over 300 time units. The parameter settings are   $g_{11}=0.5,g_{12}=1,g_{22}=-1.5,\psi_1 = \tanh(0.6x),~\psi_2=3.255\sech(4x)$.}\label{fig1}
\end{figure}

 The dark-bright soliton is governed by coupled nonlinear Schr\"{o}dinger
 equations, which can well describe  the dynamics of two-component Bose-Einstein condensates \cite{coupled1,coupled2,SoliPhy,BECreview}. The equations can be written as follows:
	
	\begin{equation}\label{xde}
		\begin{split}
		\textrm{i}\frac{\partial \psi_1}{\partial t}& = -\frac{1}{2}\frac{\partial^2 \psi_1}{\partial x^2}+(g_{11}|\psi_1|^2+g_{12}|\psi_2|^2)\psi_1,\\
		\textrm{i}\frac{\partial \psi_2}{\partial t}& = -\frac{1}{2}\frac{\partial^2 \psi_2}{\partial x^2}+(g_{12}|\psi_1|^2+g_{22}|\psi_2|^2)\psi_2,
		\end{split}
	\end{equation}
where $\psi_1$ and $\psi_2$  stand for the wave function of two-component. The nonlinear strength, $g_{11}$ ($g_{22}$) represents the interatomic interaction for the first (second) component, and $g_{12}$ is interatomic interaction between two components. The model transforms into the well-known integrable Manakov model when $g_{11}=g_{12}=g_{22}$  \cite{Manakov}. The system allows to construct the integrable exact vector soliton solutions (IES)  by the B\"{a}cklund transformation \cite{Darb1,Darb2,Darb3}, inverse scattering method \cite{inverse}, and Hirota bilinear method \cite{Hirota}, in the form bright-bright, dark-bright and dark-dark soliton \cite{vs1,BD1,BD2,BD3}. The exact solutions derived by these methods require their widths to be identical. eg. The spectrum parameter naturally determines the vector soliton has the same width when solved by the B\"{a}cklund transformation.  When integrability is destroyed, the non-integrable exact vector soliton solutions (NIES) have also been derived by periodic wave expansion method \cite{vs5,Jo} and Lagrangian variational method \cite{LVM}. These IES and NIES always require the two-component soliton solutions have the same width parameter.  We also made the widths of two-component identical in our calculations by Lagrangian variational method, as it is complex to handle different width \cite{LVM}. However, there are no physical laws dictating that the solitons in different components must have the same width. The complexity of the calculations does not imply the nonexistence of different widths solitons. Here, we derive the solutions of different width dark-bright (DWDB) soliton solutions  using Lagrangian variational method, specifically addressing the  different widths case in calculations (details see Appendix \ref{AA}). We designate the first and second component represent dark and bright soliton, respectively. The expressions for the solutions are as follows:
\begin{equation}
	\begin{split}
	\psi_1&=\textrm{i}\sqrt{a_1^2-f_1^2}+f_1\tanh[w_1(x-vt)],\\
	~~\psi_2&=f_2\sech[w_2(x-vt)]\exp(\textrm{i}vx),
	\end{split}
\end{equation}
where $f_{1,2}$ and $w_{1,2}$ stand for the amplitude and the inverse width of dark (bright) component, respectively. $a_1$ denotes the background of dark component, and $v$ is the velocity of dark-bright soliton. The constraint conditions among the parameters are provided in Appendix \ref{AA}. The stability of novel solitons are supported by plenty of numerical simulations, one of which is illustrated in Fig. \ref{fig1}. Despite a significant difference in the widths of dark and bright soliton, they remain stable during dynamical evolution, even when introducing a $3\%$ random noise in the initial states given by Lagrangian method.

\begin{figure*}[t]
	\centering
	\includegraphics[width=180mm]{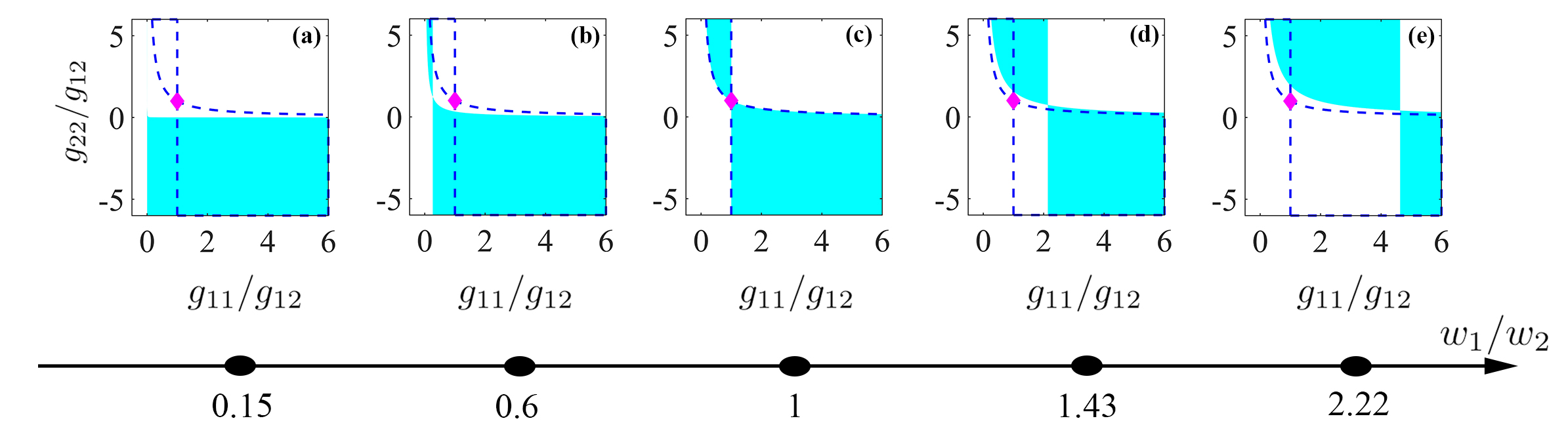}
	\caption{ The existence region (denoted by cyan area) of  DWDB soliton for various width ratios ($w_1/w_2=0.15\to2.22$) in nonlinear parameters space ($g_{12}>0$). The magenta rhombus represent the IES, while the area enclosed by blue dash line corresponding to NIES (specifically, the cyan area of (c) ). The existed scope (cyan area) of DWDB soliton family varies with width ratio, covering near entire nonlinear parameter plane when $g_{12}>0$. The different width dark-bright soliton shown in Fig. \ref{fig1} corresponds to one case of (a). }\label{fig2}
\end{figure*}

\begin{figure}[t]
	\centering
	\includegraphics[width=86mm]{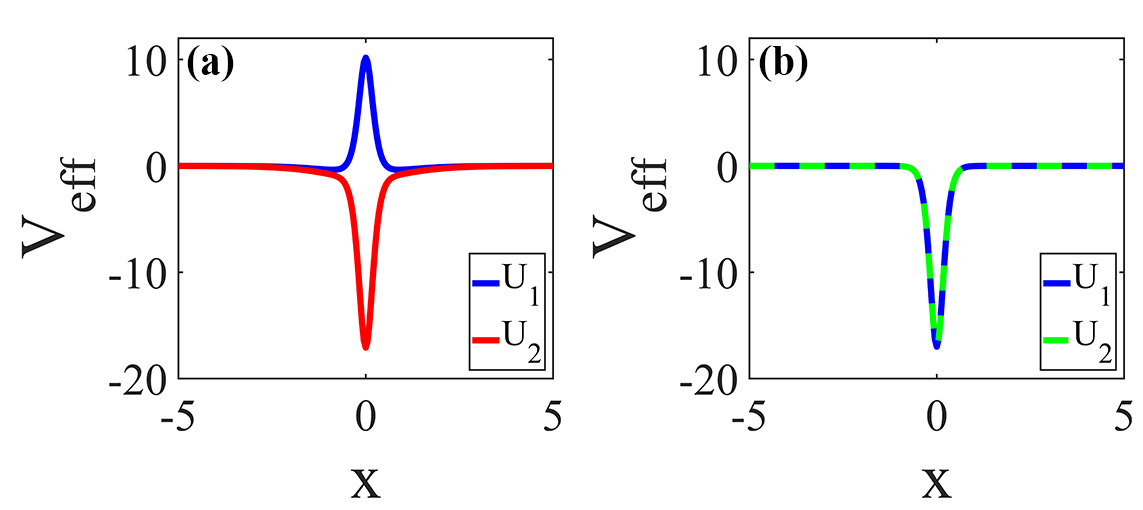}
	\caption{ (a) The effective quantum well of the first component is represented by blue line, and second component is denoted by red line in Fig. \ref{fig1}. This illustration highlights the difference in effective quantum wells for two components. (b) The effective quantum well of two components in integrable case. The quantum wells for both components consistently exhibit the same form.}\label{fig3}
\end{figure}

The DWDB solitons exhibit striking width difference between two components, distinguishing them from the previous IES and NIES. Furthermore, the DWDB solitons can exist in the regions where IES and NIES are not supported. These significantly broaden the existence scope of dark-bright soliton in nonlinear parameter space. Specifically, their existence regions depend on the ratio of two-component widths (details see Appendix \ref{AA}). Here, we choose examples with  $w_1/w_2=0.15,0.6,1,1.43,2.22$, along with $g_{12}>0$, as an illustration to depict the existence scope of DWDB soliton in nonlinear parameters space, as shown in Fig. \ref{fig2}. The magenta rhombus represents the IES of dark-bright soliton, while the area enclosed by blue dash line corresponding to NIES.  The cyan area represents the scope of DWDB soliton family, which can vary for different width ratios (as $0.15\to2.22$). The DWDB soliton degenerates to  NIES when $w_1/w_2=1$ (shown in (c)). The white area in (c) does not accommodate the IES and NIES, while the DWDB soliton solutions can exist. When adjusting width parameters, the DWDB soliton can exist in almost all regions of nonlinear parameter plane for $g_{11}>0$. In addition, these figures show that the existence scope of DWDB solutions consistently deviate the integrable case (denoted by magenta rhombus). The observation explains why the DWDB soliton are not discovered in previous experiment, where the chosen scattering lengths were in approach to the integrable case \cite{cs3,cs4}. The loss of integrability allows the dark-bright soliton to exhibit striking differences in width for many cases.

The problem that naturally arises is:  Why does the coupled non-integrable system allow for dark-bright solitons with different width parameters? We provide a possible reason why the novel soliton can exist.  Since a nonlinear Schr\"{o}diner equation can be mapped into linear Schr\"{o}diner equation with corresponding quantum well \cite{qw1,qw2,qw3}, the nonlinear term (e.g., $g_{11}|\psi_1|^2+g_{12}|\psi_2|^2$) can be understood  as an effective quantum wells. The soliton states are then associated as eigenstates within the quantum well. An example of effective quantum well for DWDB soliton are shown in Fig. \ref{fig3}(a), where $U_1= g_{11}(|\psi_1|^2-1)+g_{12}|\psi_2|^2,~U_2= g_{12}(|\psi_1|^2-1)+g_{22}|\psi_2|^2$. The two components of DWDB soliton manifest different effective quantum wells, in contrast to the previous INS where both components consistently display the same effective quantum well due to identical width (details of the proof see Appendix \ref{AB}), as shown in Fig. \ref{fig3}(b).  In an integrable system ($g_{11}=g_{12} = g_{22}$), a strong symmetry exists between the two components, predominantly admitting the identical quantum wells. However, the introduction of a non-integrable condition ($g_{11}\neq g_{12} \neq g_{22}$)  breaks the some symmetry for integrable system, allowing for the existence of different quantum wells between  two components. The presence of distinct effective quantum wells enable the existence of DWDB soliton states. The profile of effective quantum wells  and the existence region of  DWDB soliton are significantly influenced by the width ratio. Clearly, controlling the width ratio is a crucial aspect.

\subsection{Controlling the width ratio between two components}

\begin{figure}[t]
	\centering
	\includegraphics[width=88mm]{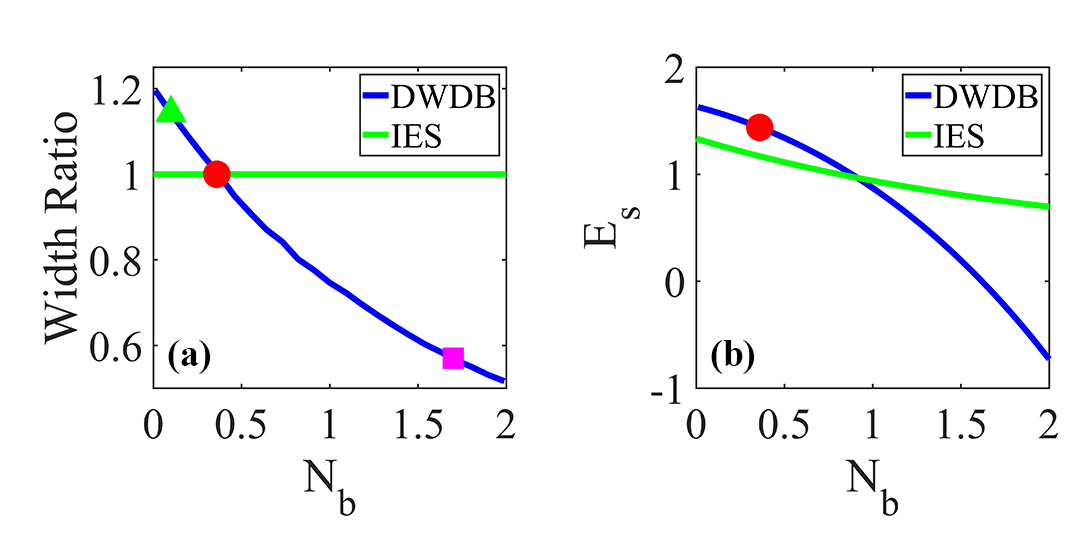}
	\caption{(a) The relation between the widths ratio $w_1/w_2$ and $N_b$ is examined for two solutions: different widths dark-bright solitons (DWDB) and the integrable exact soliton  solutions (IES). For DWDB: The width ratios decrease with an increase in $N_b$. When the width ratio $w_1/w_2$ is greater than $1$ (less than $1$), the dark-component's width is narrower (wider) than that of bright component is observed. This is because $w_{1,2}$  represent inverse widths. For instance,  a green triangle (a magenta square) denote $w_1/w_2>1$ ($w_1/w_2<1$). The red point represent a special case of DWDB ($w_1/w_2=1$), equivalent to NIES.  IES: The width of two-component is consistently identical for varies of $N_b$. (b) The excitation energy versus $N_b$ is examined for the two solutions: DWDB and IES (detailed in Appendix \ref{AC}), with the red dot indicating NIES.  The parameter settings are $g_{11}=1.5,g_{12}=1,g_{22}=-1.5,v=0$ for DWDB; $ g_{11}=g_{12}=g_{22}=1,v=0$ for IES.}\label{fig4}
\end{figure}

The DWDB soliton solutions, detailed in Appendix \ref{AA}, involve two arbitrary free parameters. Since the widths of two components are too narrow to  properly observe in experiment, and are difficult to directly control them. We aim to modify the soliton width ratio by the controllable variables. In experimental setups, the generation of dark-bright soliton in experiment are with the aid of spatial-dependent Rabi coupling and phase imprinting techniques \cite{cs3}, which control the particle number of bright soliton (denoted as $N_b=\int_{-\infty}^{+\infty}|\psi_2|^2\textrm{d}x$) and velocity ($v$), respectively. Therefore, in the following discussion, our focus is the influence of system due to the two variables (the new expressions see Appendix \ref{AC}). The relation between the widths ratio $w_1/w_2$  for  dark-bright solitons and $N_b$, while keeping the nonlinear parameters $g_{11},g_{12},g_{22}$ and $v$ fixed,  as shown in Fig. \ref{fig4}(a).  For the DWDB, the width ratios decrease with an increase in $N_b$. When the width ratio $w_1/w_2$ is greater than $1$ (less than $1$), the dark-component's width is narrower (wider) than that of bright component is observed, as $w_{1,2}$  represent inverse widths of soliton. These indicate that the width ratio of DWDB soliton can be effectively controlled  by  fixing the value of $v$ and manipulating the particle number of bright soliton  $N_b$.  While for the IES, they consistently exhibit the same width across varying $N_b$. In Fig. \ref{fig4}(b), we illustrate the excitation energy of DWDB soliton and IES, respectively. In contrast to the integrable case, the excitation energy of DWDB soliton obviously decrease with an increase of $N_b$ even can exhibit negative excitation energy. This implies that the energy required for stimulating DWDB solitons is less than that needed for IES. In addition to having lower excitation energy, the DWDB solutions are abundant across diverse nonlinear parameters and $N_b$. This suggests that the DWDB soliton may be readily generated in experimental conditions where the scattering length deviates from integrable case. Subsequently, we conduct  numerical simulations to generate the DWDB solitons based on current experiment technology \cite{cs3}.

\section{The generation of different widths dark-bright soliton}\label{iii}	

	\begin{figure}[t]
	\centering
	\includegraphics[width=89mm]{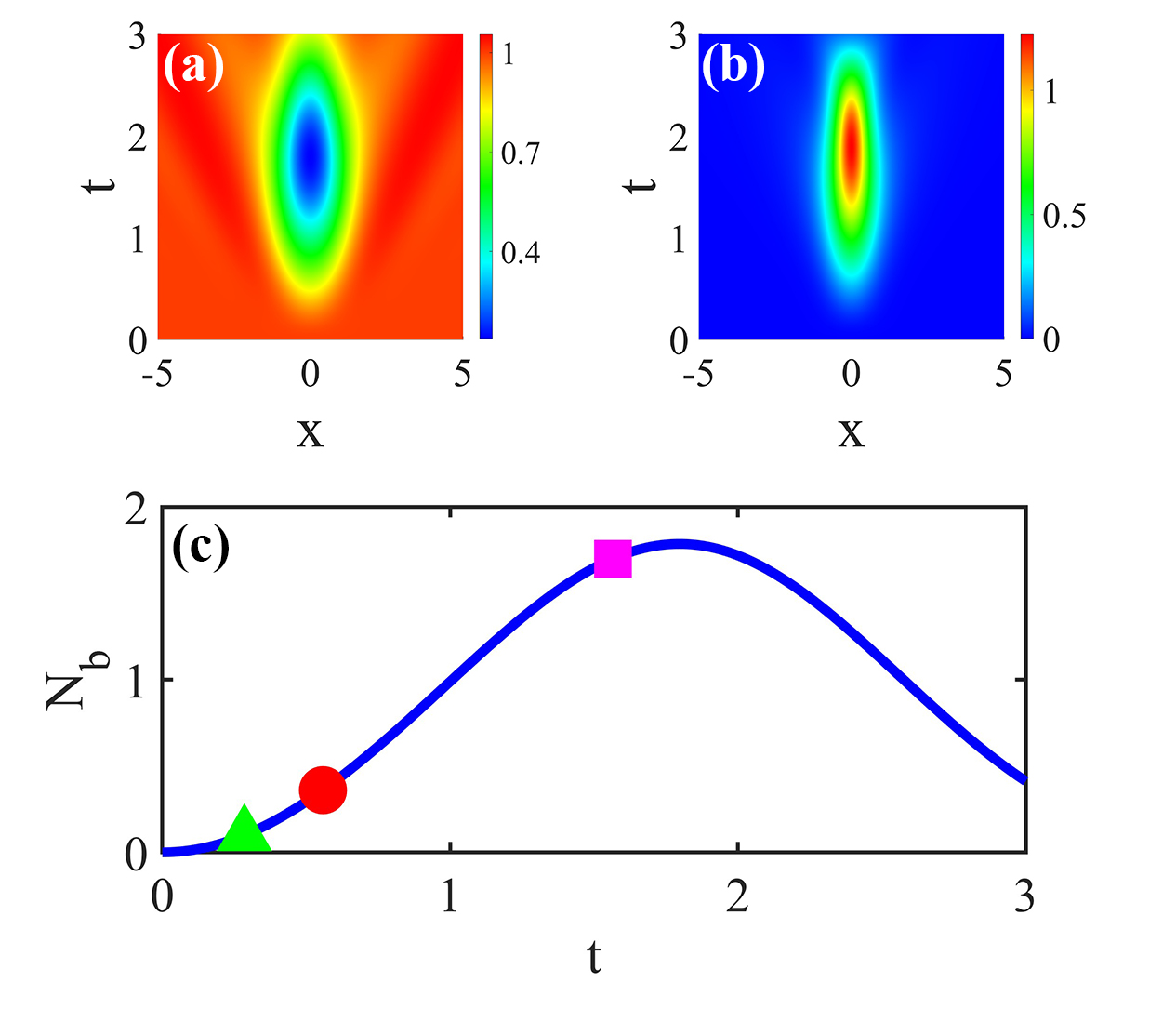}
	\caption{ (a)(b) The numerical evolution with one cycle for the initial states which the first (second) component is plane wave $\psi_1=1$ (none $\psi_2=0$). (c) The number of atom for bright soliton $N_b$ versus time evolution obtained by numerical statistic. The red dot denotes the atom numbers for which needed by the NIES ($N_b=0.36$) in the case. The green triangle and magenta square represent the atom numbers required for two DWDB solitons ($N_b = 0.1,t _1= 0.29$ and $N_b = 1.7,t_2=1.566$ ), respectively.}\label{fig5}
\end{figure}

\begin{figure*}[t]
	\centering
	\includegraphics[width=180mm]{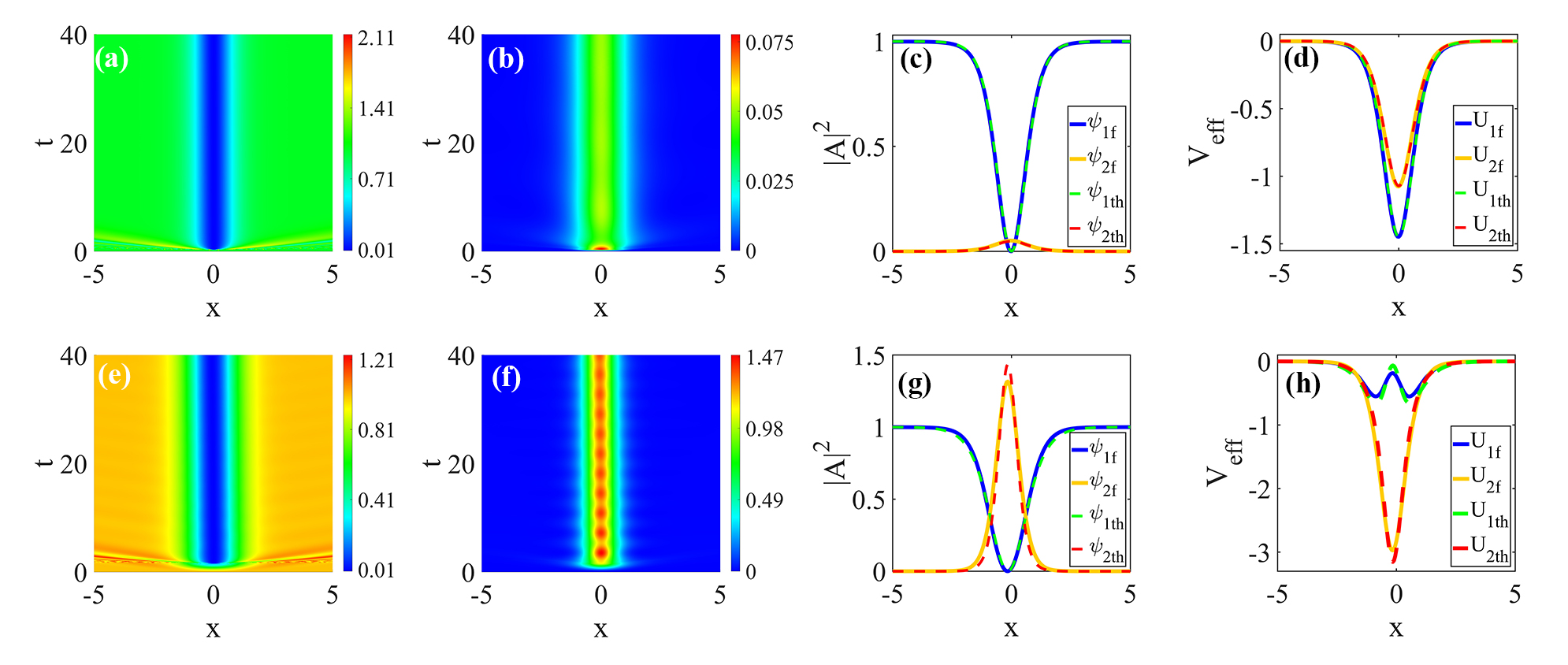}
	\caption{ (a)(b) and (e)(f) The density evolution in numerical simulations is obtained by closing the atom transition ($\Omega=0$) in the time $t_1=0.29$ ( denoted by green triangle from Fig. \ref{fig5}) and $t_2=1.566$ (denoted by magenta square from Fig. \ref{fig5}). Simultaneously, we imprint the phase $\phi~(p\to1)$ in first component.  (c)(g) The profile of two types DWDB soliton, with the solid line denoting the final profile of numerical result, while dashed line denotes  the theoretical result. (d)(h) The effective quantum well of two types DWDB soliton. The theoretical and numerical results agree well with each other. The parameter settings are $g_{11}=1.5,g_{12}=1,g_{22}=-1.5,\Omega(x) = \exp(-x^2)$, (a)(b)(c)(d) $t_1=0.29,N_b=0.1,v\to0,\psi_{1th} =\tanh(1.2x),\psi_{2th}=0.2187\sech(1.05x)$. (e)(f)(g)(h) $t_2 =1.566,N_b=1.7,v\to0,\psi_{1th} =\tanh[0.96(x+0.15)],\psi_{2th}=1.1985\sech[1.96(x+0.15)]$. As the velocity is not strictly zero, the final state deviates the initial position. We perform spatial translation for the theoretical results from Appendix \ref{AC} to facilitate comparison with numerical results. }\label{fig6}
\end{figure*}	

We have analyzed the DWDB soliton solutions in the preceding section, our results indicate that the DWDB soliton widely exists in non-integrable case ($g_{11}\neq g_{12}\neq g_{22}$), and different width ratios can be effectively controlled  by  fixing the value of $v$ and manipulating the particle number of bright soliton $N_b$.  Investigating how to generate them based on this characteristic in an experimental setup presents an intriguing challenge. Considering that the previously generate the dark-bright soliton was successfully achieved through the combination of optical transfer (phase-locked Raman laser) and  phase imprinting technology in ${^{87}}$Rb Bose-Einstein condensates \cite{cs3}. The designed numerical simulation scheme is: In the first step, we adjust atomic nonlinear interactions away from integrable case. In the second step, we generate the ground state (i.e., a plane wave denoted as $\psi_1=1$) in the first component, while the second component is initially empty, represent by $\psi_2=0$. In the third step, atoms are transformed from the first component to second component  through local  atom transition. Finally, after a certain duration, we generate the dark soliton by imprint phase $\phi=\arctan[p\tanh(px)/\sqrt{1-p^2}]+\frac{\pi}{2}$ on the first component. Simultaneously, with the closing of the atom transition. The dark soliton in the first component creates a potential well that traps atoms in the second component by inter-species nonlinear interaction, resulting in the generation of the DWDB soliton. The number of  atom for the bright soliton and the system velocity are controlled by the timing of the atom transition and the phase of imprint, respectively. It is noteworthy that the generation of DWDB soliton only requires adjusting atomic nonlinear interactions away from the integrable case based on the experimental technology \cite{cs3}.  We provide the two examples to illustrate the scheme intuitively. Considering a quasi-one-dimensional two-component Bose-Einstein condensates
with transition effect, the dynamics of system can be described
by a dimensionless coupled nonlinear Schr\"{o}diner equations with Rabi coupling \cite{RCS1,RCS2,RCS3}:
	\begin{eqnarray}
			\textrm{i}\frac{\partial \psi_1}{\partial t}\!&=&\! -\frac{1}{2}\frac{\partial^2 \psi_1}{\partial x^2}+(g_{11}|\psi_1|^2+g_{12}|\psi_2|^2)\psi_1+\Omega(x)\psi_2,\nonumber\\\\
			\textrm{i}\frac{\partial \psi_2}{\partial t}\!&=\!& -\frac{1}{2}\frac{\partial^2 \psi_2}{\partial x^2}+(g_{12}|\psi_1|^2+g_{22}|\psi_2|^2)\psi_2+\Omega(x)\psi_1, \nonumber
	\end{eqnarray}
where $\Omega(x) = a\exp[-\frac{(x-c)^2}{b}]$ which can lead local atom transition. Here we set $a=1,c=0,b=1$, $\psi_1=1,\psi_2=0,g_{11}=1.5,g_{12}=1,g_{22}=-1.5$. The system exhibits period local atom transition in the initial condition, with the numerical evolution of one cycle shown in Fig. \ref{fig5}(a)(b). We obtain the number of atoms $N_b$ for the second component versus time evolution through numerical statistic, as shown in Fig. \ref{fig5}(c). This indicates that we can control $N_b$ by manipulating the transition time, thereby control the width ratio of DWDB solitons. We choose width ratios $w_1/w_2$ are $1.143$ and $0.57$ as examples to generate the DWDB solitons. The corresponding particle numbers $N_b$  are $0.1$ and $1.7$ denoted by a green triangle and a magenta square in Fig. \ref{fig4}, respectively. We conduct two numerical simulations, separately closing the atom transition ($\Omega=0$) at times $t_1=0.29$ and $t_2=1.566$ to obtain the corresponding $N_b=0.1$ and $N_b=1.7$. Simultaneously, we imprint the phase $\phi~(p\to1)$ in the first component. The whole evolution process is shown in Fig. \ref{fig6}(a)(b) and (e)(f). The DWDB solitons with different width ratios are generated successfully. The theoretical results $\psi_{1th},\psi_{2th}$  agree well with the final state of numerical results $\psi_{1f},\psi_{2f}$, as shown in \ref{fig6}(c)(g). The effective potential well shown in Fig. \ref{fig6}(d)(h). The examples exhibit the DWDB soliton can be generated easily in numerical simulation, and they agree well with our theoretical results. We expect that the generation of DWDB solitons in experiments is feasible.  Referring to \cite{cs3}, the experimental project is designed as follows:

\subsection*{Experimental Scheme}	

We produce a BEC composed of $5\times 10^4$ $^{87}$Rb atoms in the
$5^2S_{1/2}$ , $\ket{F = 1, m_F = -1}$ state in an optical dipole trap with trapping frequencies $\omega_z = 2\pi\times5.9 $Hz, $\omega^
{ver}_{\perp} = 2\pi \times 85$ Hz and $\omega^
{hor}_{\perp} = 2\pi\times133 $Hz. The classical scattering lengths are $ a_{11} = 100.86a_0, a_{12}=
98.98a_0$, and $a_{22} = 94.57a_0$, where $a_0 = 5.29 \times 10^{-11}$ m is the Bohr radius, $a_{11,12,22}$ are direct proportion to $g_{11,12,22}$. Firstly, we adjust the scattering lengths by Feshbach resonances technology \cite{FR,FR2}, ensuring they deviate from the integrable case. This involves setting parameters like $ a_{11} >0, a_{12}>0$ and $a_{22}<0$. This adjustment is crucial because the presence of DWDB soliton is widespread in the specified region (refer to Fig. \ref{fig2}). Secondly, the condensate consists of atoms in the
$\ket{F=1,m_F=-1}$ state is transferred to  $\ket{F=2,m_F=0}$ state, which is accomplished by the use of a phase-locked Raman laser system with a relative phase. Thirdly, the application of the phase imprinting light pattern via the spatial light modulator for a certain duration, the generation of a dark soliton is induced. The dark soliton creates an effective potential well in the
$\ket{F=1,m_F=-1}$ state, trapping atoms form a bright soliton in the $\ket{F=2,m_F=0}$ state by inter-species nonlinear interaction. This phenomenon occurs because the bright soliton is ground sate of the effective potential well. Finally, we switch off the optical transfer within a period of time,  the DWDB soliton is successfully generated. The number of  atom for bright soliton and system velocity are controlled by the action time of the optical transfer and the phase imprinting, respectively. The generation of DWDB solitons only requires changing the nonlinear interaction via Feshbach resonances based on the previous experimental technology used for dark-bright soliton generation.
	
\section{conclusion and discussion}\label{iv}
We obtain the novel dark-bright soliton solution family with striking width difference by Lagrangian variational method, and these solitons remain stable when the random noise is introduced in initial states. The different width dark-bright (DWDB)  soliton exist extensively in nonlinear parameter space, significantly broadening the range of the dark-bright soliton solutions compared the integrable and non-integrable exact solutions. The width ratios of DWDB  soliton can be effectively controlled  by  fixing the value of $v$ and manipulating the particle number of bright soliton $N_b$. And the width ratio can deviate from $1$ when the nonlinear interaction moves away from the integrable case.  Therefore, we suggest that the generation of DWDB solitons can be realised by changing the nonlinear interaction via Feshbach resonances based on the previous experimental technology used for dark-bright soliton generation. The results indicate that abundant vector solitons with difference widths exist in multi-components coupled systems, and would motivate experiments to obverse the novel vector solitons in coupled nonlinear system, such as Bose-Einstein condensates \cite{BEC}, nonlinear optical fibers \cite{fiber1,fiber2}. Additionally, our results can be utilized to derive  solutions for different width bright-bright, dark-dark soliton and other multi-component soliton. The various characteristics of different width solitons need further exploration in the future, such as the dispersion relation and inertial mass for different solitons' widths.

	\section*{Acknowledgments}
	This work was supported by the National Natural Science Foundation of China (Contract No. 12235007, 12375005,12022513),  and the Major Basic Research Program of Natural Science of Shaanxi Province (Grant No. 2018KJXX-094).

\begin{widetext}
\begin{appendix}
			
\section{the derivation of different width dark-bright soliton and their existence region}\label{AA}
We consider two-component coupled nonlinear Schr\"{o}dinger equation, which can be written as	
		\begin{equation}\label{A1}
			\begin{split}
				\textrm{i}\frac{\partial \psi_1}{\partial t}& = -\frac{1}{2}\frac{\partial^2 \psi_1}{\partial x^2}+(g_{11}|\psi_1|^2+g_{12}|\psi_2|^2)\psi_1,\\
				\textrm{i}\frac{\partial \psi_2}{\partial t}& = -\frac{1}{2}\frac{\partial^2 \psi_2}{\partial x^2}+(g_{12}|\psi_1|^2+g_{22}|\psi_2|^2)\psi_2,
			\end{split}
		\end{equation}	
where $\psi_{1,2}$ stand for the wave function of two components. The nonlinear parameters $g_{11,22}$ ($g_{12}$) is the intra-species interaction for the first or second component (inter-species interaction between two components). We derive the different width dark-bright  (DWDB) soliton solutions by the Lagrangian
variational method. We introduce the Lagrangian density
as $\mathcal{L} =\frac{\textrm{i}}{2}(\psi_1^*\partial_t\psi_1-\psi_1\partial_t\psi_1^*)
(1-\frac{a_1^2}{|\psi_1|^2})-\frac{1}{2}|\partial_x\psi_1|^2 -\frac{g_{11}}{2}(|\psi_1|^2-a_1^2)^2+\frac{\textrm{i}}{2}(\psi_2^*\partial_t\psi_2-\psi_2\partial_t\psi_2^*)-\frac{1}{2}|\partial_x\psi_2|^2-\frac{g_{22}}{2}|\psi_2|^2-g_{12}(|\psi_1|^2-a_1^2)|\psi_2|^2$. We adopt the trial function of DWDB soliton, as
	\begin{eqnarray}\label{shitan}
				\psi_{1D}=\!\!\!\!&&\textrm{i}\sqrt{a_1^2-f_1^2(t)}+f_1(t)\tanh\{w_1(t)[x-b(t)]\},\nonumber\\\\
				\psi_{2B}=\!\!\!\!&&f_2(t)\sech\{w_2(t)[x-b(t)]\}\textrm{e}^{\textrm{i}\{\xi(t)+b'(t)[x-b(t)]\}},\nonumber
	\end{eqnarray}
	where $\psi_1$ ($\psi_2$) denotes the wave function of the dark (bright) soliton, $f_{1,2}$ and $w_{1,2}$ describe the amplitude and width of the dark (bright) soliton, respectively, $a_1$ is the background of the dark component, the central position of the soliton is $b(t)$, $\xi$ is the time-dependent phases of the bright components. We obtain the Lagrangian ($L$) by substituting Eq. \eqref{shitan} into $\mathcal{L}$ and integrating over space from $-\infty$ to $+\infty$, which can be simplified as
	\begin{equation}
		\begin{split}
			L=&-2f_1(t)\sqrt{a_1^2-f_1^2(t)}b'(t)+2 a_1^2\arcsin\big[\frac{f_1(t)}{a_1}\big]b'(t)-\frac{2}{3}f_1^2(t)w_1(t)-\frac{2}{3}g_{11}\frac{f_1^4(t)}{w_1(t)}\\&+f_2^2(t)\frac{b'^2(t)}{w_2(t)}-2f_2^2(t)\frac{\xi_2'(t)}{w_2(t)}-\frac{1}{3}f_2^2(t)w_2(t)-\frac{2}{3}g_{22}\frac{f_2^4(t)}{w_2(t)}
			-g_{12}f_1^2(t)f_2^2(t)G(w_1(t),w_2(t)),
		\end{split}
 \end{equation}		
			 where $	G =-\int^{+\infty}_{-\infty}\sech^2\{w_1(t)[x-b(t)]\} \sech^2\{w_2(t)[x-b(t)]\}\textrm{d}x $.  The  motion equations of system are obtained by applying the Euler-Lagrangian equations, and the non-trivial equations are:	$-g_{11}f_1^2+\frac{3}{2}g_{12}f_2^2w_1^2\frac{\partial G}{\partial w_1}+w_1^2=0, ~~	g_{22}f_2^2+\frac{3}{2}g_{12}f_1^2w_2G+\frac{3}{2}g_{12}f_1^2w_2^2\frac{\partial G}{\partial w_2}+w_2^2=0,~~
			 (2g_{11}f_1^2+\frac{3}{2}g_{12}f_2^2w_1G+w_1^2)\sqrt{a_1^2-f_1^2}-3f_1w_1b'(t)=0$. The DWDB soliton solutions can be simplified as
			 \begin{equation}\label{A4}
			 		\psi_1=\textrm{i}\sqrt{a_1^2-f_1^2}+f_1\tanh[w_1(x-vt)],
			 		~~~~~~~~~\psi_2=f_2\sech[w_2(x-vt)]\exp(\textrm{i}vx),
			 \end{equation}			
				where $b'=v, ~b(t) = vt,~f_1=\sqrt{\frac{g_{22}-g_{12}H_1}{g_{11}g_{22}-g_{12}^2H_3}} w_1, ~~f_2 = \sqrt{\frac{g_{12}H_2-g_{11}}{g_{11}g_{22}-g_{12}^2H_3}}w_2,~~v = \frac{\sqrt{a_1^2-f_1^2}}{6f_1w_1}(4f_1^2g_{11}+3f_2^2Gg_{12}w_1+2w_1^2),$ $ H_1=\frac{3}{2}w_2^2\frac{\partial G}{\partial w_1},~H_2 =- \frac{3}{2} w_1^2\big(\frac{G}{w_2}+\frac{\partial G}{\partial w_2}\big),~H_3=-\frac{9}{4}w_1^2w_2\frac{\partial G}{\partial w_1}\big(G+w_2\frac{\partial G}{\partial w_2}\big)$. The existence of DWDB soliton requires the nonlinear coefficients satisfy: 	$\frac{1}{g_{12}}(\frac{g_{22}}{g_{12}}-H_1)>0,~\frac{g_{11}}{g_{12}}\frac{g_{22}}{g_{12}}-H_3>0,~\frac{1}{g_{12}}(H_2-\frac{g_{11}}{g_{12}})>0$ (or all expression are negative). Their existence region relies on width of two components, which degenerate to non-integrable exact solution when $w_1=w_2$.  Our following discussion is based on numerical integral result of $G$ because its complexity. The contour lines of the value for $H_1,H_2,H_3$ in two-component width parameters $w_1,w_2$ space are shown in Fig. \ref{figA1}. The contour lines are straight lines in the $w_1,w_2$ space, indicating that the values of $H_1,H_2, H_3$ depend solely on the ratio of the two-component widths ($w_1/w_2$). Therefore, we only focus on $w_1/w_2$ when discussing the existence region for DWDB soliton in nonlinear parameters space, as shown in Fig. \ref{figA2}. The existence region for DWDB soliton changes with varying $w_1/w_2$. These characters broaden extensively existence region of DWDB soliton, compared with the integrable and non-integrable exact solutions. The DWDB solitons exhibit good stability in dynamical evolution when $g_{12}>0$, while they usually show instability when $g_{12}<0$. The systematic study of stability  needs further work.
			
		\begin{figure}[t]
			\centering
			\includegraphics[width=180mm]{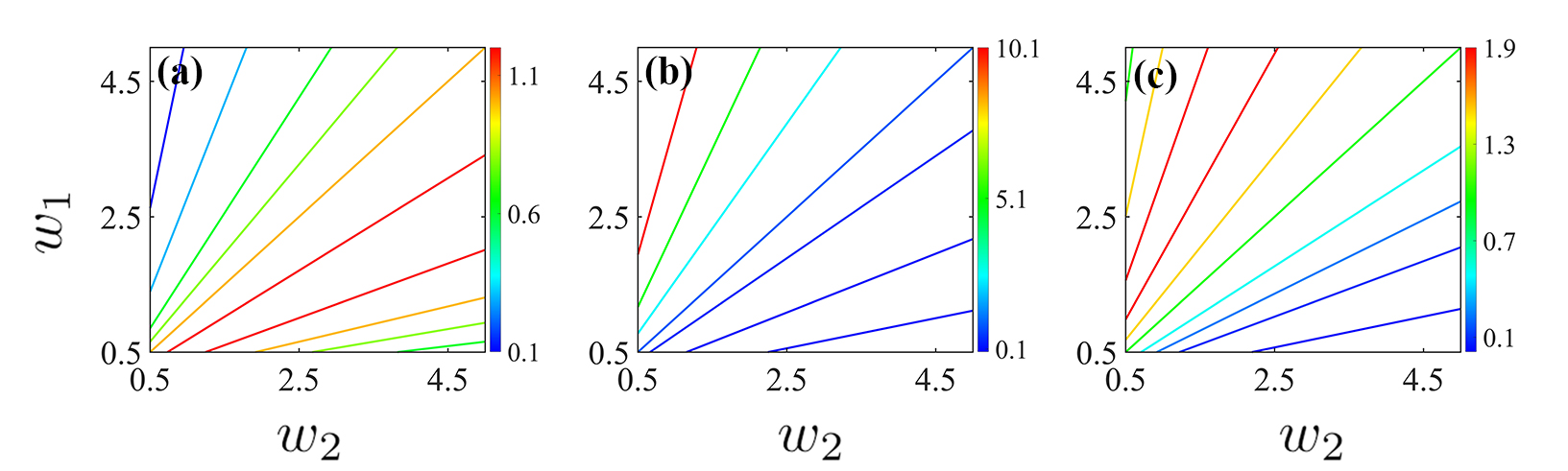}
			\caption{ (a)(b)(c) The contour lines of the values for $H_1,H_2$ and $H_3$ in two-component width parameters $w_1,w_2$ space, respectively. The colorbar represent the values of $H_1,H_2, H_3$. The contour lines are straight lines in the $w_1,w_2$ space, indicating that the values of $H_1,H_2, H_3$  depend solely on the ratio of two-component widths ($w_1/w_2$).}\label{figA1}
		\end{figure}
	
				\begin{figure}[t]
				\centering
				\includegraphics[width=160mm]{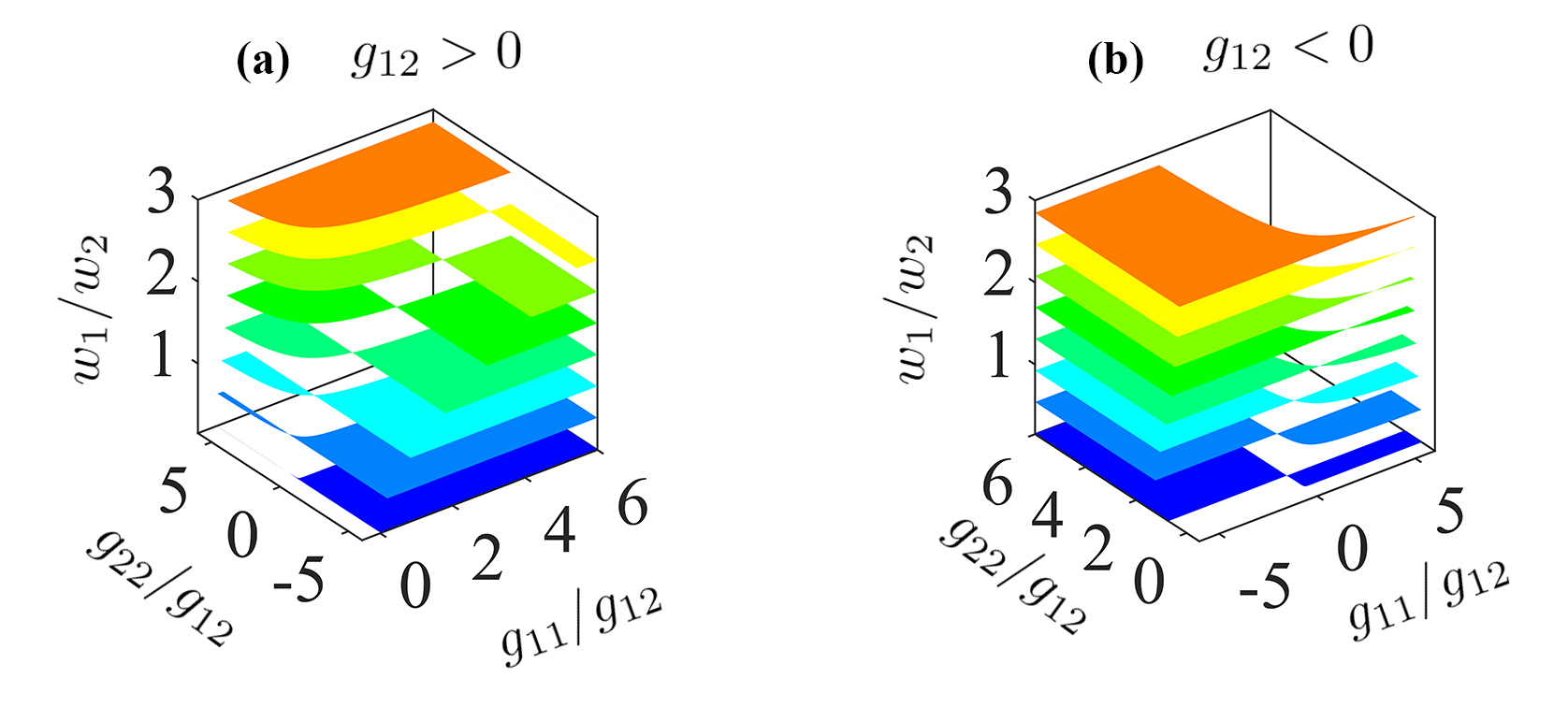}
				\caption{ (a) The existence region of different width dark-bright soliton in nonlinear parameters space for $g_{12}>0$. The colormap denotes the ratio of two-component width parameters $w_1/w_2$. The different width soliton can almost exist in the entire nonlinear parameter plane ($g_{11}/g_{12}>0$) by varying width parameters. (b) The existence region of different width dark-bright soliton in nonlinear parameters space for $g_{12}<0$. }\label{figA2}
			\end{figure}

			\section{The effective quantum well of the exact dark-bright soliton solutions}\label{AB}
		The expression of dark-bright soliton  solution is \cite{LVM}
		\begin{equation}
			\begin{split}
				\psi_{1}&=\big\{\textrm{i}\sqrt{a_1^2-f_1^2}+f_1\tanh[w(x-vt)]\big\}\textrm{e}^{-{\rm{i}}g_{11}a_1^2t},\\
				\psi_{2}&=f_2\sech[w(x-vt)]\textrm{e}^{\textrm{i}[\frac{1}{2}(w^2+v^2)t+(x-vt)v]-\textrm{i}g_{12}a_1^2t}.
			\end{split}
		\end{equation}
	For the integrable condition, parameters satisfy $g_{11}=g_{12}=g_{22}=g,~w=\sqrt{gf_1^2-gf_2^2},~v=\frac{w}{f_1}\sqrt{a_1^2-f_1^2}$. The effective quantum well of integrable exact solution (IES) is
	\begin{equation}
		U_1=U_2= g(|\psi_1|^2-a_1^2)+g|\psi_2|^2=-gf_1^2\sech^2[w(x-vt)]+gf_2^2\sech^2[w(x-vt)]=-w^2\sech^2[w(x-vt)],
	\end{equation}
	For the non-integrable condition, exact solution with identical width, as a special case of a different widths dark-bright soliton, the parameters satisfy $f_2=\sqrt{\frac{g_{11}-g_{12}}{g_{12}-g_{22}}}f_1,~w=\sqrt{\frac{g_{12}^2-g_{11}g_{22}}{g_{12}-g_{22}}}f_1,~v=\frac{w}{f_1}\sqrt{a_1^2-f_1^2}$. The effective quantum well of non-integrable exact solution (NIES) is
			\begin{equation}
			\begin{split}
				U_1&= g_{11}(|\psi_1|^2-1)+g_{12}|\psi_2|^2=-g_{11}f_1^2\sech^2[w(x-vt)]+g_{12}f_2^2\sech^2[w(x-vt)]=-w^2\sech^2[w(x-vt)],\\
				U_2&= g_{12}(|\psi_1|^2-1)+g_{22}|\psi_2|^2=-g_{12}f_1^2\sech^2[w(x-vt)]+g_{22}f_2^2\sech^2[w(x-vt)]=-w^2\sech^2[w(x-vt)],
			\end{split}
			\end{equation}
			In conclusion, the effective quantum well of the two components always has the same form for INS and NIES with identical width.
			
		\section{Controlling the width ratio between two components}\label{AC}		
		We concentrate on the two variables of system: the particle number of bright soliton $N_b$ (denoted by $\int_{-\infty}^{+\infty}|\psi_2 |^2dx$) and the velocity $v$. The Eq. \eqref{A4} can be expressed as
			\begin{equation}
				\psi_1=\textrm{i}\sqrt{a_1^2-f_1^2}+f_1\tanh[w_1(x-vt)],
				~~~~~~~~~\psi_2=\sqrt{\frac{N_bw_2}{2}}\sech[w_2(x-vt)]\exp(\textrm{i}vx),
			\end{equation}
			where $f_1=\sqrt{\frac{g_{22}-g_{12}H_1}{g_{11}g_{22}-g_{12}^2H_3}} w_1, ~~N_b = 2\frac{g_{12}H_2-g_{11}}{g_{11}g_{22}-g_{12}^2H_3}w_2,~~v = \frac{\sqrt{a_1^2-f_1^2}}{6f_1w_1}(4f_1^2g_{11}+3f_2^2Gg_{12}w_1+2w_1^2),$ $ H_1=\frac{3}{2}w_2^2\frac{\partial G}{\partial w_1},~H_2 =- \frac{3}{2} w_1^2\big(\frac{G}{w_2}+\frac{\partial G}{\partial w_2}\big),~H_3=-\frac{9}{4}w_1^2w_2\frac{\partial G}{\partial w_1}\big(G+w_2\frac{\partial G}{\partial w_2}\big),G =-\int^{+\infty}_{-\infty}\sech^2[w_1(x-vt)]\sech^2[w_2(x-vt)]\textrm{d}x $.
			
We investigate their influence on system by varying $N_b$ while keeping the velocity $v$ fixed. For simplicity, we set $f_1=a_1=1,v=0$, the constraint condition can be simplified as
\begin{equation}\label{C2}
	\sqrt{\frac{g_{22}-g_{12}H_1}{g_{11}g_{22}-g_{12}^2H_3}} w_1=1, ~~~ 2\frac{g_{12}H_2-g_{11}}{g_{11}g_{22}-g_{12}^2H_3}w_2=N_b.
\end{equation}
The widths ratio $w_1/w_2$ of system can be solved when the value of $N_b$ is given. As an example, we set $g_{11}=1.5,g_{12}=1,g_{22}=-1.5$, the relation between the widths ratio $w_1/w_2$ and $N_b$ shown in Fig. \ref{fig4}(a). Successfully controlling the system's width ratio $w_1/w_2$ is achieved by adjusting $N_b$ while maintaining a constant velocity $v$. Additionally, the particle number $N_b$ also affects the excitation energy of system. For instance, setting $g_{11} = 1.5,g_{12}=1,g_{22}=-1.5$,  the relation between $E_s$ and $N_b$ shown in Fig. \ref{fig4}(b). $E_s=\int_{-\infty}^{+\infty}\frac{1}{2}|\partial_x\psi_1|^2 +\frac{g_{11}}{2}(|\psi_1|^2-a_1^2)^2+\frac{1}{2}|\partial_x\psi_2|^2+\frac{g_{22}}{2}|\psi_2|^2+g_{12}(|\psi_1|^2-a_1^2)|\psi_2|^2dx=\frac{2}{3}f_1^2w_1  +\frac{2}{3}g_{11}\frac{f_1^4  }{w_1  }+\frac{N_b}{2}v^2 +\frac{1}{6}N_b  w_2^2  +\frac{1}{6}g_{22}N_b^2w_2
+\frac{1}{2}g_{12}f_1^2  N_bw_2 G$. Two examples of DWDB solitons, the parameters setting is $g_{11}=1.5,g_{12}=1,g_{22}=-1.5,v=0$. From Eq. \eqref{C2}, $N_b=0.1,\psi_1 =\tanh(1.2x),\psi_2=0.2187\sech(1.05x)$; $N_b=1.7,\psi_1 =\tanh(0.96x),\psi_2=1.1985\sech(1.96x)$.

		\end{appendix}
	\end{widetext}


\begin{thebibliography}{99}
		\bibitem{SoliPhy}  L. Pitaevskii and S. Stringari, \textit{Bose-Einstein condensation} (Oxford University Press, Oxford, England, 2003).
		
		\bibitem{BEC}  P. G. Kevrekidis, D. Frantzeskakis, and R. Carretero-Gonz\'{a}lez, \textit{Emergent Nonlinear Phenomena in Bose-Einstein Condensates: Theory and Experiment} (Springer,
		Berlin Heidelberg, 2009).
		
		\bibitem{OS}   J. R. Taylor, \textit{Optical Solitons---Theory and Experiment} (Cambridge University Press, Cambridge, 1992).
		
		\bibitem{cs1}  S. Burger, K. Bongs, S. Dettmer, W. Ertmer, K. Sengstock, A. Sanpera, G. V. Shlyapnikov, and M. Lewenstein, Dark Solitons in Bose-Einstein Condensates, \newblock
		\href{https://doi.org/10.1103/PhysRevLett.83.5198} {Phys. Rev. Lett. \textbf{83}, 5198 (1999)}.
		
		\bibitem{cs2}  L. Khaykovich, F. Schreck, G. Ferrari, T. Bourdel,
		J. Cubizolles, L. D. Carr, Y. Castin and C. Salomon, Formation of a Matter-Wave
		Bright Soliton, \newblock
		\href{https://doi.org/10.1126/science.1071021} {Science \textbf{296}, 5571 (2002)}.
		
		\bibitem{cs3}  C. Becker, S. Stellmer, P. S. Panahi, S. D\"{o}rscher, M. Baumert, E. M. Richter, J. Kronj\"{a}ger, K. Bongs and K. Sengstock, Oscillations and interactions of dark and dark-bright solitons in Bose-Einstein condensates, \newblock
		\href{https://doi.org/10.1038/nPhys962} {Nat. Phys. \textbf{4}, 496  (2008)}.
		
		\bibitem{cs4}  C. Hamner, J. J. Chang, P. Engels, and M. A. Hoefer, Generation of Dark-Bright Soliton Trains in Superfluid-Superfluid Counterflow, \newblock\href{https://doi.org/10.1103/PhysRevLett.106.065302} {Phys. Rev. Lett. \textbf{106}, 065302  (2011)}.
		
	
		\bibitem{cs5}  E. Kengne, W.-M. Liu, and B. A. Malomed, Spatiotemporal engineering of matter-wave solitons
		in Bose-Einstein condensates, \newblock\href{https://doi.org/10.1016/j.physrep.2020.11.001} {Phys. Rep. \textbf{899}, 1  (2021)}.
		

		\bibitem{cs6}  J. Ieda, T. Miyakawa, and M. Wadati, Exact Analysis of Soliton Dynamics in Spinor Bose-Einstein Condensates, \newblock\href{https://doi.org/10.1103/PhysRevLett.93.194102} {Phys. Rev. Lett. \textbf{93}, 194102 (2004)}.
		
		\bibitem{cs7} C. Qu, L. P. Pitaevskii, and S. Stringari, Magnetic Solitons in a Binary Bose-Einstein Condensate, \newblock\href{https://doi.org/10.1103/PhysRevLett.116.160402} {Phys. Rev. Lett. \textbf{116}, 160402  (2016)}.
		
		\bibitem{cs8}  X. Yu and P. B.  Blakie, Propagating Ferrodark Solitons in a Superfluid: Exact Solutions and Anomalous Dynamics, \newblock\href{https://doi.org/10.1103/PhysRevLett.128.125301} {Phys. Rev. Lett. \textbf{128}, 125301  (2022)}.
		
		
		
		\bibitem{vs1}  D. J. Kaup and B. A. Malomed, Soliton trapping and daughter waves in the Manakov model, \newblock\href{https://doi.org/10.1103/PhysRevA.48.599} {Phys. Rev. A \textbf{48}, 599  (1993)}.
		
		\bibitem{vs2}  X. Liu, H. Pu, B. Xiong, W. M. Liu, and J. Gong, Formation and transformation of vector solitons in two-species Bose-Einstein condensates with a tunable interaction, \newblock\href{https://doi.org/10.1103/PhysRevA.79.013423} {Phys. Rev. A \textbf{79}, 013423  (2009)}.
		
		\bibitem{vs3}  T. M. Bersano, V. Gokhroo, M. A. Khamehchi, J. D'Ambroise, D. J. Frantzeskakis, P. Engels, and P. G. Kevrekidis, Three-Component Soliton States in Spinor
		F=1 Bose-Einstein Condensates, \newblock
		\href{https://doi.org/10.1103/PhysRevLett.120.063202} {Phys. Rev. Lett. \textbf{120}, 063202 (2018)}.
		
		\bibitem{vs4}  A. Farolfi, D. Trypogeorgos, C. Mordini, G. Lamporesi, and G. Ferrari, Observation of Magnetic Solitons in Two-Component Bose-Einstein Condensates, \newblock
		\href{https://doi.org/10.1103/PhysRevLett.125.030401} {Phys. Rev. Lett. \textbf{125}, 030401 (2020)}.
		
		\bibitem{vs5}  X. Chai, D. Lao, K. Fujimoto, R. Hamazaki, M. Ueda, and C. Raman, Magnetic Solitons in a Spin-1 Bose-Einstein Condensate, \newblock
		\href{https://doi.org/10.1103/PhysRevLett.125.030402} {Phys. Rev. Lett. \textbf{125}, 030402 (2020)}.
		
	
		\bibitem{vs6}  L.-Z. Meng, Y.-H. Qin and  L.-C. Zhao, Spin solitons in spin-1 Bose-Einstein condensates, \newblock\href{https://doi.org/10.1016/j.cnsns.2022.106286} {Commun. Nonlinear Sci. Numer. Simulat. \textbf{109}, 106286  (2022)}.
			
	
		\bibitem{inverse} V. E. Zakharov and A. B. Shabat, Exact theory of two-dimensional self-focusing and one-dimensional self-modulation of waves in nonlinear media, Sov. Phys. JETP \textbf{34}, 62
	(1972)
	
	
	\bibitem{AS1} D. H. Peregrine, Water waves, nonlinear Schr\"odinger equations and their solutions,
	\newblock\href{https://doi.org/10.1017/S0334270000003891} {J. Aust. Math. Soc. Ser. B \textbf{25}, 16 (1983).}
	
	\bibitem{AS2} Y. S. Kivshar, V. V. Afansjev, and A. W. Snyder, Dark-like bright solitons, \newblock
	\href{https://doi.org/10.1016/0030-4018(96)00111-3} {Opt. Commun. \textbf{126}, 348 (1996).}
	
		\bibitem{KDS}  B. Luther-Davies and Y. S. Kivshar, Dark optical solitons: physics and applications, \newblock
	\href{https://doi.org/10.1016/S0370-1573(97)00073-2} {Phys. Rep. \textbf{298}, 81 (1998)}.
	
		
		\bibitem{Darb1} V. B. Matveev and M. A. Salle, \textit{Darboux Transformation and Solitons} (Springer, Berlin, 1991).
		
		\bibitem{Darb2} M. Ma\~{n}as, Darboux transformations for the nonlinear Schr\"{o}dinger equations, \newblock
		\href{http://iopscience.iop.org/0305-4470/29/23/029} {J. Phys. A \textbf{29}, 7721  (1996)}.
		
		\bibitem{Darb3}  L.-M. Ling, L.-C. Zhao, and B.-L. Guo, Darboux transformation and multi-dark
		soliton for N-component nonlinear
		Schr\"{o}dinger equations, \newblock
		\href{http://dx.doi.org/10.1088/0951-7715/28/9/3243} {Nonlinearity \textbf{28}, 3243  (2015)}.
		

		
		\bibitem{Hirota} R. Hirota, \textit{The Direct Method in Soliton Theory} (Cambridge University Press, Cambridge, 2004).
		
		\bibitem{Manakov} S. V. Manakov, On the theory of two-dimensional stationary self-focusing of electromagnetic waves, Sov. Phys.-JETP \textbf{38}, 248 (1974).
		
		\bibitem{Jo}  S.-F. Yao, Q.-Y. Li and Z.-D. Li, Combined periodic wave and solitary wave solutions in two-component Bose-Einstein condensates, \newblock
		\href{https://iopscience.iop.org/article/10.1088/1674-1056/20/11/110307} {Chin. Phys. B \textbf{20}, 110307  (2011)}.
			
		\bibitem{LVM}  N. Mao and L.-C. Zhao, Exact analytical soliton solutions of
		N-component coupled nonlinear Schr\"{o}dinger equations with arbitrary nonlinear parameters, \newblock\href{https://doi.org/10.1103/PhysRevE.106.064206} {Phys. Rev. E \textbf{106}, 064206  (2022)}.
		
		\bibitem{DBO}  Th. Busch and J. R. Anglin, Dark-Bright Solitons in Inhomogeneous Bose-Einstein Condensates, \newblock
		\href{https://doi.org/10.1103/PhysRevLett.87.010401} {Phys. Rev. Lett. \textbf{87}, 010401 (2001)}.
		
		\bibitem{multi-dark-bright}  D. Yan, J. J. Chang, C. Hamner, P. G. Kevrekidis, P. Engels, V. Achilleos, D. J. Frantzeskakis, R. Carretero-Gonz\'{a}lez, and P. Schmelcher, Multiple dark-bright solitons in atomic Bose-Einstein condensates, \newblock\href{https://doi.org/10.1103/PhysRevA.84.053630} {Phys. Rev. A \textbf{84}, 053630  (2011)}.
		
		\bibitem{DBT}  V. Achilleos, D. Yan, P. G. Kevrekidis, and D. J. Frantzeskakis, Dark-bright solitons in Bose-Einstein condensates at finite temperatures, \newblock\href{https://doi.org/10.1088/1367-2630/14/5/055006} {New J. Phys. \textbf{14}, 055006  (2012)}.
		
		
		\bibitem{DBBC}  G. C. Katsimiga, P. G. Kevrekidis, B. Prinari, G. Biondini, and P. Schmelcher, Dark-bright soliton pairs: Bifurcations and collisions, \newblock\href{https://doi.org/10.1103/PhysRevA.97.043623} {Phys. Rev. A \textbf{97}, 043623  (2018)}.
		
		\bibitem{Spin AC}  L.-C. Zhao, W. Wang, Q. Tang, Z.-Y. Yang, W.-L. Yang and J. Liu, Spin soliton with a negative-positive mass transition, \newblock
		\href{https://doi.org/10.1103/PhysRevA.101.043621} {Phys. Rev. A \textbf{101}, 043621  (2020)}.
		
		\bibitem{AC Oscillation}  S. Bresolin, A. Roy, G. Ferrari, A. Recati, and N. Pavloff, Oscillating Solitons and ac Josephson Effect in Ferromagnetic Bose-Bose Mixtures, \newblock\href{https://doi.org/10.1103/PhysRevLett.130.220403} {Phys. Rev. Lett. \textbf{130}, 220403  (2023)}.
		
		\bibitem{VM1}  D. Anderson, Variational approach to nonlinear pulse propagation in optical fibers, \newblock
		\href{https://doi.org/10.1103/PhysRevA.27.3135} {Phys. Rev. A \textbf{27}, 3135  (1983)}.
		
		\bibitem{VM2}  T. Ueda and W. L. Lath, Dynamics of coupled solitons in nonlinear optical fibers, \newblock
		\href{https://doi.org/10.1103/PhysRevA.42.563} {Phys. Rev. A \textbf{42}, 563  (1990)}.
		
		\bibitem{VM3}  D. J. Kaup, B. A. Malomed, and R. S. Tasgal, Internal dynamics of a vector soliton in a nonlinear optical fiber, \newblock
		\href{https://doi.org/10.1103/PhysRevE.48.3049} {Phys. Rev. E \textbf{48}, 3049  (1993)}.
		
		\bibitem{VM4}  Yu. S. Kivshar and W. Kr\'{o}likowski, Lagrangian approach for dark solitons, \newblock
		\href{https://doi.org/10.1016/0030-4018(94)00644-A} {Opt. Commun. \textbf{114}, 353 (1995)}.	
		
		\bibitem{VM5}  C. Par\'{e}, Accurate variational approach for vector solitary waves, \newblock
		\href{https://doi.org/10.1103/PhysRevE.54.846} {Phys. Rev. E \textbf{54}, 846  (1996)}.
		
		\bibitem{VM6}  M. O. D. Alotaibi and L. D. Carr, Dynamics of dark-bright vector solitons in Bose-Einstein condensates, \newblock
		\href{https://doi.org/10.1103/PhysRevA.96.013601} {Phys. Rev. A \textbf{96}, 013601  (2017)}.
		
		 \bibitem{qw1}  L. D. Landau and E. M. Lifshitz, \textit{Quantum Mechanics}, (Nauka, Moscow, 1989).
		
		\bibitem{qw2}  N. Akhmediev and A. Ankiewicz, Partially Coherent Solitons on a Finite Background, \newblock\href{https://doi.org/10.1103/PhysRevLett.82.2661} {Phys. Rev. Lett. \textbf{82}, 2661  (1999)}.
		
		\bibitem{qw3}  L.-C. Zhao, Z.-Y. Yang, and W.-L. Yang, Solitons in nonlinear systems and eigen-states in quantum wells, \newblock\href{https://iopscience.iop.org/article/10.1088/1674-1056/28/1/010501} {Chin. Phy. B \textbf{28}, 010501  (2019)}.
		
		\bibitem{coupled1} F. T. Hioe, Solitary Waves for $N$ Coupled Nonlinear Schr\"{o}dinger Equations, \newblock
		\href{https://doi.org/10.1103/PhysRevLett.82.1152} {Phys. Rev. Lett. \textbf{82}, 1152 (1999)}.
		
		\bibitem{coupled2}  T. Dauxois and M. Peyrard, \textit{Physics of Solitons} (Cambridge University, Cambridge, 2006).
		
		\bibitem{BECreview}  F. Dalfovo, S. Giorgini, L. P. Pitaevskii, S. Stringari, Theory of Bose-Einstein condensation in trapped gases, \newblock\href{https://doi.org/10.1103/RevModPhys.71.463} {Rev. Mod. Phys.  \textbf{71}, 463  (1999)}.
		
		
		\bibitem{BD1}  R. Radhakrishnan and M. Lakshmanan, Bright and dark soliton solutions to coupled nonlinear
		Schr\"{o}dinger equations, \newblock
		\href{http://iopscience.iop.org/0305-4470/28/9/025} {J. Phys. A \textbf{28}, 2683  (1995)}.
		
		\bibitem{BD2}  A. P. Sheppard and Yu. S. Kivshar, Polarized dark solitons in isotropic Kerr media, \newblock
		\href{https://doi.org/10.1103/PhysRevE.55.4773} {Phys. Rev. E \textbf{55}, 4773  (1997)}.
		
		\bibitem{BD3}  Y.-H. Qin, Y. Wu, Z.-Y. Yang and L.-C. Zhao, Interference properties of two-component matter wave solitons, \newblock
		\href{https://iopscience.iop.org/article/10.1088/1674-1056/ab65b7} {Chin. Phys. B \textbf{29}, 020303  (2020)}.
		

		
		\bibitem{RCS1}  Q.-Han Park and J. H. Eberly, Strong Confinement and Oscillations in Two-Component Bose-Einstein Condensates, \newblock\href{https://doi.org/10.1103/PhysRevLett.85.4195} {Phys. Rev. Lett. \textbf{85}, 4195  (2000)}.
		
		\bibitem{RCS2}  C. Qu, M. Tylutki, S. Stringari, and L. P. Pitaevskii, Magnetic solitons in Rabi-coupled Bose-Einstein condensates, \newblock\href{https://doi.org/10.1103/PhysRevA.95.033614} {Phys. Rev. A \textbf{95}, 033614  (2017)}.
		
		\bibitem{RCS3}  L.-C. Zhao, G.-G. Xin, and Z.-Y. Yang, Transition dynamics of a bright soliton in a binary Bose-Einstein condensate, \newblock\href{https://doi.org/10.1364/JOSAB.34.002569} {J.
			Opt. Soc. Am. B \textbf{34}, 2569  (2017)}.
		
		\bibitem{FR}  S. Inouye, M. Andrews, J. Stenger, H.-J. Miesner, D. M.Stamper-Kurn, and W. Ketterle, Observation of Feshbach resonances in a Bose-Einstein condensate, \newblock\href{https://doi.org/10.1038/32354} {Nature (London) \textbf{392}, 151  (1998)}.
		
		\bibitem{FR2}  S. Roy, M. Landini, A. Trenkwalder, G. Semeghini, G.	Spagnolli, A. Simoni, M. Fattori, M. Inguscio, and G.
		Modugno, Test of the Universality of the Three-Body Efimov Parameter at Narrow Feshbach Resonances, \newblock\href{https://doi.org/10.1103/PhysRevLett.111.053202} {Phys. Rev. Lett. \textbf{111}, 053202  (2013)}.
		
		
		\bibitem{fiber1}  C. Sulem and P. L. Sulem, \textit{Nonlinear Schr\"{o}dinger Equations:	Self-Focusing Instability and Wave Collapse} (Springer, New York, 1999).
		
		\bibitem{fiber2} G. P. Agrawal, \textit{Nonlinear Fiber Optics} (Academic Press, New York,1995).
		
		
		
		
	\end{thebibliography}
\end{document}